\documentstyle[12pt,prd,aps,epsfig,floats,preprint,eqsecnum]{revtex}
\tightenlines

\newcommand{\be}{\begin{equation}}
\newcommand{\ee}{\end{equation}}
\newcommand{\bea}{\begin{eqnarray}}
\newcommand{\eea}{\end{eqnarray}}
\newcommand{\no}{\nonumber}
\def\lsim{\mathrel{\rlap{\lower4pt\hbox{\hskip1pt$\sim$}}
    \raise1pt\hbox{$<$}}}         
\def\gsim{\mathrel{\rlap{\lower4pt\hbox{\hskip1pt$\sim$}}
    \raise1pt\hbox{$>$}}}         
\begin{document}
\begin{titlepage}
\preprint{
\vbox{   \hbox{CERN-TH/2001-125}
         \hbox{IFIC/01-23}
         \hbox{WIS/9/01-May-DPP}
         \hbox{hep-ph/0105159}}}

\title{New CP Violation in Neutrino Oscillations}
\author{M. C. Gonzalez-Garcia$^{1,2,3}$ \thanks{concha@thwgs.cern.ch},
Y. Grossman$^{4}$ \thanks{yuvalg@physics.technion.ac.il},
A. Gusso$^{1,5}$ \thanks{gusso@ific.uv.es} and
Y. Nir$^6$ \thanks{ftnir@wicc.weizmann.ac.il}} 
\vskip 1cm
\address{$^1$Instituto de F\'{\i}sica Corpuscular,  
Universitat de  Val\`encia -- C.S.I.C\\
Edificio Institutos de Paterna, Apt 22085, 46071 Val\`encia, Spain\\
$^2$ Theory Division, CERN CH1211, Geneva 23, Switzerland \\
$^3$ C.N. Yang Institute for Theoretical Physics\\
State University of New York at Stony Brook\\
Stony Broow,NY 11794-3840, USA\\
$^4$ Department of Physics, Technion--Israel Institute of Technology\\
Technion City, 32000 Haifa, Israel\\
$^5$ Instituto de F\'{\i}sica Te\'orica, Universidade Estadual Paulista\\
     Rua Pamplona 145, 01405--900, S\~ao Paulo, Brazil \\
$^6$ Department of Particle Physics, Weizmann Institute of Science\\
 Rehovot 76100, Israel} 
\maketitle
\begin{abstract}
Measurements of CP--violating observables in neutrino oscillation
experiments have been studied in the literature as a way to determine
the CP--violating phase in the mixing matrix for leptons. Here we show
that such observables also probe new neutrino interactions in the
production or detection processes. Genuine CP violation and fake CP
violation due to matter effects are sensitive to the imaginary and
real parts of new couplings.  The dependence of the CP asymmetry on
source--detector distance is different from the standard one and, in
particular, enhanced at short distances.  We estimate that future
neutrino factories will be able to probe in this way new interactions
that are up to four orders of magnitude weaker than the weak
interactions. We discuss the possible implications for models of new
physics.
 
\end{abstract}

\end{titlepage}

\section{New CP Violation in Neutrino Interactions}
In the future, neutrino oscillation experiments will search for 
CP--violating effects 
\cite{DeRujula:1999hd,Barger:2000fs,Tanimoto:1999tj,Minakata:1999ze,dghr,Romanino:2000zq,Koike:2000hf,Freund:2000gy,cerv,Mocioiu:2000st,Barger:2000yn,Minakata:2000ee,Kimura:2000kb,Yokomakura:2000sv,Bueno:2000jy,Parke:2001hu,Lipari:2001ds,Burguet-Castell:2001ez,Freund:2001pn,Akhmedov:2001kd,Freund:2001ui,Pinney:2001xw}. 
The Standard Model, extended to include masses for the light, active neutrinos, 
predicts that CP is violated in neutrino oscillations through a single phase in 
the mixing matrix for leptons. This effect is suppressed by small mixing angles 
and small mass differences.

It is not unlikely, however, that the high-energy physics that is
responsible for neutrino masses and mixing involves also new neutrino
interactions. Such interactions provide new sources of CP
violation. In this work we study CP--violating effects due to
contributions from new neutrino interactions to the production and/or
detection processes in neutrino oscillation experiments. We
investigate the following questions:

(i) How would new, CP--violating neutrino interactions manifest
themselves in neutrino oscillations?

(ii) Are the effects qualitatively different from the Standard Models ones?
In particular, can we use the time (or, equivalently, distance) dependence 
of the transition probability to distinguish between Standard Model and new
CP violation?

(iii) How large can the effects be? In particular, do the new interactions
suffer from suppression factors related to mixing angles and mass differences?

(iv) Can the new CP violation be observed in proposed experiments? What would
be the optimal setting for these observations?

(v) Which models of New Physics can be probed in this way?

The plan of this paper goes as follows. In section II we present a
parameterization of the New Physics effects that are of interest to us
and explain the counting of independent CP--violating phases in our
framework.  In section III we evaluate the New Physics effects on the
transition probability in neutrino vacuum oscillation experiments. (A
full expression for the transition probability, without any
approximations concerning mixing angles and mass differences, is given
in Appendix A.) In section IV we investigate the resulting CP
asymmetry and compare the New Physics contribution to the standard one
(that is, the contribution to the asymmetry from lepton mixing). In
sections V and VI we evaluate the New Physics effects on, respectively,
the transition probability and CP asymmetry, in neutrino matter
oscillations. In section VII we study how these effects can be
observed in future neutrino factory experiments. In particular, we
estimate a lower bound on the strength of the new interactions that
can be observed in these experiments. This lower bound is compared to
existing model--independent upper bounds in section VIII.  We
summarize our results and discuss some of the implications that
would arise if a signal is experimentally observed in section IX.
 
\section{Notations and Formalism}
In this section we give a model--independent parameterization of 
New Physics
effects on production and detection processes in neutrino oscillation
experiments. We put special emphasis on CP--violating phases.

We denote by $|\nu_i\rangle$, $i=1,2,3$, the three neutrino mass eigenstates.
We denote by $|\nu_\alpha\rangle$ the weak interaction partners of the
charged lepton mass eigenstates $\alpha^-$ ($\alpha=e,\mu,\tau$):
\be
\label{nuellW}
|\nu_\alpha\rangle=\sum_i U_{\alpha i}|\nu_i\rangle.
\ee

Whenever we use an explicit parameterization of the lepton mixing 
matrix \cite{MNS,koma}, we will use the most conventional one:
\be
\label{MNSpar}
U\equiv U_{23}U_{13}U_{12}\equiv
\pmatrix{1&0&0\cr 0&c_{23}&s_{23}\cr 0&-s_{23}&c_{23}\cr}
\pmatrix{c_{13}&0&s_{13}e^{i\delta}\cr 0&1&0\cr -s_{13}e^{-i\delta}&0&c_{13}\cr}
\pmatrix{c_{12}&s_{12}&0\cr -s_{12}&c_{12}&0\cr 0&0&1\cr},
\ee
with $s_{ij}\equiv\sin\theta_{ij}$ and $c_{ij}\equiv\cos\theta_{ij}$.
Alternatively, a convention-independent definition of the phase $\delta$
that we will use in our calculations is given by
\be
\label{dkmdef}
\delta\equiv\arg\left({U_{e3}U_{\mu3}^*\over U_{e1}U_{\mu1}^*}\right).
\ee

We consider new, possibly CP--violating, physics in the production and/or 
detection process. Such effects were previously studied in Ref.~\cite{Yuval},
and we follow closely the formalism of that paper. Most of the analysis in
\cite{Yuval}, however, was carried out assuming CP conservation.
We parameterize the New Physics interaction in the source 
and in the detector by two sets of effective four--fermion couplings, 
$(G^s_{\rm NP})_{\alpha\beta}$ and $(G^d_{\rm NP})_{\alpha\beta}$,
where $\alpha,\beta=e,\mu,\tau$. Here $(G^s_{\rm NP})_{\alpha\beta}$
refers to processes in the source where a $\nu_\beta$ is produced
in conjunction with an incoming $\alpha^-$ or an outgoing $\alpha^+$
charged lepton, while $(G^d_{\rm NP})_{\alpha\beta}$
refers to processes in the detector where an incoming $\nu_\beta$ 
produces an $\alpha^-$ charged lepton. While the $SU(2)_L$ gauge
symmetry requires that the four--fermion couplings of the charged current
weak interactions be proportional to $G_F\delta_{\alpha\beta}$, new
interactions allow couplings with $\alpha\neq\beta$. Phenomenological
constraints imply that the new interaction is suppressed with respect to 
the weak interaction,
\be
\label{smallnp}|(G_{\rm NP}^s)_{\alpha\beta}|\ll G_F,\ \ \ 
|(G_{\rm NP}^d)_{\alpha\beta}|\ll G_F.
\ee

For the sake of concreteness, we consider the production and detection
processes that are relevant to neutrino factories. We therefore study
an appearance experiment where neutrinos are produced in the process
$\mu^+\to e^+\nu_\alpha\bar\nu_\mu$ and detected by the
process $\nu_\beta d\to\mu^- u$, and antineutrinos are produced and
detected by the corresponding charge-conjugate processes.  Our results
can be modified to any other neutrino oscillation experiment in a
straightforward way. The relevant couplings are then $(G_{\rm
NP}^s)_{e\beta}$ and $(G_{\rm NP}^d)_{\mu\beta}$. It is convenient to
define small dimensionless quantities $\epsilon^{s,d}_{\alpha\beta}$
in the following way:
\bea
\label{defeps}
\epsilon^s_{e\beta}&\equiv& {(G_{\rm NP}^s)_{e\beta}\over
\sqrt{|G_F+(G_{\rm NP}^s)_{ee}|^2+|(G_{\rm NP}^s)_{e\mu}|^2
+|(G_{\rm NP}^s)_{e\tau}|^2}},\no\\
\epsilon^d_{\mu\beta}&\equiv& {(G_{\rm NP}^d)_{\mu\beta}\over
\sqrt{|G_F+(G_{\rm NP}^d)_{\mu\mu}|^2+|(G_{\rm NP}^d)_{\mu e}|^2
+|(G_{\rm NP}^d)_{\mu\tau}|^2}}.
\eea
Since we assume that $|\epsilon^{s,d}_{\alpha\beta}|\ll1$, we will only
evaluate their effects to leading (linear) order. New flavor--conserving
interactions affect neutrino oscillations only at ${\cal O}(|\epsilon|^2)$
and will be neglected from here on. (More precisely, the leading effects from 
flavor--diagonal couplings are proportional to 
$\epsilon$ (flavor--diagonal)$\times\epsilon$(flavor--changing)  
and can therefore be safely neglected.)

We use an explicit parameterization for only two of the $\epsilon$'s,
with the following convention:
\be
\label{deleps}
\epsilon^s_{e\mu}\equiv|\epsilon^s_{e\mu}|e^{i\delta_\epsilon},\ \ \ \ 
\epsilon^{d*}_{\mu e}\equiv|\epsilon^{d*}_{\mu e}|e^{i\delta^\prime_\epsilon}.
\ee
Alternatively, we can define the phases $\delta_\epsilon$ and
$\delta^\prime_\epsilon$ in a convention-independent way:
\be
\label{defdeps}
\delta_\epsilon\equiv\arg\left({\epsilon^s_{e\mu}\over U_{e1}U_{\mu1}^*}\right),
\ \ \ \ 
\delta^\prime_\epsilon\equiv\arg\left({\epsilon^{d*}_{\mu e}\over 
U_{e1}U_{\mu1}^*}\right).
\ee

We would like to conclude this section with a comment on the number of
independent CP--violating phases in our framework. It is well known that the 
three--generation mixing matrix for leptons depends, in the case of Majorana
neutrinos, on three phases. Two of these, related to the fact that there is
no freedom in redefining the phases of neutrino fields, do not affect
neutrino oscillations and are therefore irrelevant to our discussion.
The other one is analogous to the Kobayashi--Maskawa phase of the mixing
matrix for quarks. The freedom of redefining the phases of charged lepton 
fields is fully used to reduce the number of relevant phases to one.
Consequently, it is impossible to remove any phases from the
$\epsilon^{s,d}_{\alpha\beta}$ parameters. Each of these parameters 
introduces a new, independent CP--violating phase. 

\begin{figure}
\centerline{\mbox{\epsfig{file=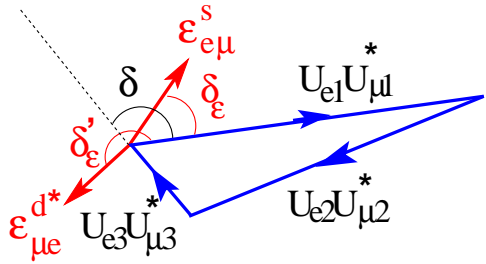,width=0.6\textwidth,angle=-90}}}
\caption{
The neutrino parameters that dominate $P_{e\mu}$ in the complex plane. 
We show the relevant unitarity triangle, which is the geometrical presentation
of the relation $U_{e1}U_{\mu1}^*+U_{e2}U_{\mu2}^*+U_{e3}U_{\mu3}^*=0$, and the 
two parameters that describe the New Physics in the production, 
$\epsilon^s_{e\mu}$, and in the detector, $\epsilon^{d*}_{\mu e}$. The three 
independent phases defined in the text, $\delta$, $\delta_\epsilon$ and 
$\delta^\prime_\epsilon$, are shown explicitly. The standard convention puts 
$U_{e1}U_{\mu1}^*$ on the real axis.}
\label{trieps}
\end{figure}

For example, when we discuss $\nu_e\to\nu_\mu$ oscillations, our results
will depend on $\epsilon^s_{e\mu}$, $\epsilon^d_{\mu e}$ and the 
$U_{ei}U_{\mu i}^*$ ($i=1,2,3$) mixing parameters. This set of parameters
depends on three independent phases, one of which is the 
$\delta$ of Eq.~(\ref{dkmdef}), while the other two can be chosen to be 
$\delta_\epsilon$ 
and $\delta^\prime_\epsilon$ of Eq.~(\ref{defdeps}). This situation is 
illustrated in Fig. \ref{trieps}, where we show in the complex plane the 
unitarity triangle and the $\epsilon^{s,d}$ parameters that 
are most relevant to $\nu_e\to\nu_\mu$ oscillations.

\section{The Transition Probability in Vacuum} 
In this section we derive the expression for the transition probability
in neutrino oscillation experiments as a function of the mixing matrix
parameters and the New Physics parameters.

We denote by $\nu_e^s$ the neutrino state that is produced in the source in
conjunction with an $e^+$, and by $\nu_\mu^d$ the neutrino state that is 
signalled by $\mu^-$ production in the detector:
\bea
\label{nusd}
|\nu_e^s\rangle&=&\sum_i\left[U_{e i}+\epsilon^s_{e\mu}U_{\mu i}
+\epsilon^s_{e\tau}U_{\tau i}\right]|\nu_i\rangle,
\no\\
|\nu_\mu^d\rangle&=&\sum_i\left[U_{\mu i}+\epsilon^d_{\mu e}U_{e i}
+\epsilon^d_{\mu\tau}U_{\tau i}\right]|\nu_i\rangle.
\eea
(Note that the norm of the states so defined is one up to effects of
${\cal O}(|\epsilon|^2)$, which we consistently neglect.)
We obtain the following expression for the transition probability
$P_{e\mu}=|\langle\nu_\mu^d|\nu_e^s(t)\rangle|^2$, where $\nu_e^s(t)$
is the time-evolved state that was purely $\nu_e^s$ at time $t=0$ :
\be
\label{probem}
P_{e\mu}=\left|\sum_i e^{-iE_it}\left[U_{e i}U^*_{\mu i}
+\epsilon^s_{e\mu}|U_{\mu i}|^2+\epsilon^{d*}_{\mu e}|U_{e i}|^2
+\epsilon^s_{e\tau}U_{\tau i}U^*_{\mu i}
+\epsilon^{d*}_{\mu\tau}U_{\mu i}U^*_{\tau i}\right]\right|^2.
\ee
Our results will be given in terms of $\Delta m^2_{ij}$, $\Delta_{ij}$
and $x_{ij}$, which are defined as follows:
\be
\Delta m^2_{ij}\equiv m_i^2-m_j^2,\ \ \ 
\Delta_{ij}\equiv\Delta m^2_{ij}/(2E),\ \ \ x_{ij}\equiv\Delta_{ij}L/2,
\label{defDelx}
\ee
where $E$ is the neutrino energy and $L$ is the distance between 
the source and the detector.

Equation (\ref{probem}) will be the starting point of our calculations.
The full expression for $P_{e\mu}$ in vacuum is given in Appendix A
and has been used for our numerical calculations described below. To
understand the essential features of our analysis it is, however, more
useful to do the following. First, we separate $P_{e\mu}$ into a
Standard Model piece, $P_{e\mu}^{\rm SM}$, and a New Physics piece,
$P_{e\mu}^{\rm NP}$. What we mean by $P_{e\mu}^{\rm SM}$ is
$P_{e\mu}(\epsilon^{s,d}_{\alpha\beta}=0)$. This is the contribution
to $P_{e\mu}$ from the Standard Model extended to include neutrino
masses but no new interactions. In contrast, $P_{e\mu}^{\rm NP}$
contains all the $\epsilon^{s,d}_{\alpha\beta}$--dependent
terms. Second, since the atmospheric and reactor neutrino data imply
that $|U_{e3}|$ is small and the solar neutrino data imply that
$\Delta m^2_{12}/\Delta m^2_{13}$ is small, we expand $P_{e\mu}^{\rm
SM}$ to second order and $P_{e\mu}^{\rm NP}$ to first order in
$|U_{e3}|$ and $\Delta m^2_{12}$.  

For $P^{\rm SM}_{e\mu}$ we obtain:
\bea
\label{pemsm}
P_{e\mu}^{\rm SM}& = & 4x_{21}^2 |U_{e2}|^2 |U_{\mu 2}|^2 +
4\sin^2 x_{31} |U_{e3}|^2 |U_{\mu 3}|^2 \no\\
&+&4 x_{21}\sin 2x_{31} 
{\cal R}e \bigr(U_{e2}U^*_{e3} U^*_{\mu 2} U_{\mu 3}\bigl)\no\\
&-&8x_{21}\sin^2x_{31}{\cal I}m\bigr(U_{e2}U^*_{e3} U^*_{\mu 2} U_{\mu 3}\bigl).
\eea
The first term is the well known transition probability in the 
two--generation case. The second term gives the well known transition
probability in the approximation that $\Delta m^2_{12}=0$. The last
term is a manifestation of the Standard Model CP violation.

For $P_{e\mu}^{\rm NP}$ we obtain
\bea
\label{pemnp}
P_{e\mu}^{\rm NP}=&-&4 \sin^2 x_{31}{\cal R}e\biggr[U_{e3}^*U_{\mu 3}
\left(\epsilon^{d*}_{\mu e}+\epsilon^s_{e\mu} (1-2|U_{\mu3}|^2)
-2 \epsilon_{e\tau}^s U^*_{\mu 3} U_{\tau 3}\right)\biggl]\no\\
&+&4 x_{21} \sin 2 x_{31}{\cal R}e\biggl[U_{e2}^*U_{\mu 2}
\left(\epsilon_{e\mu}^s|U_{\mu 3}|^2
+\epsilon^s_{e\tau} U_{\mu 3}^* U_{\tau 3}\right)\biggr]\no\\
&-&4 x_{21}{\cal I}m\biggl[U_{e2}^* U_{\mu 2}\left(\epsilon^{d*}_{\mu e}
+\epsilon_{e\mu}^s (1-|U_{\mu 3}|^2)
-\epsilon_{e\tau}^s U_{\mu 3}^* U_{\tau 3}\right)\biggr]  \no \\
&-& 2 \sin 2x_{31}{\cal I}m\biggl[ U_{e3}^*U_{\mu 3}
(\epsilon^{d*}_{\mu e}+\epsilon^s_{e\mu})\biggr]\no\\
&-&4 x_{21}\cos2 x_{31}  
{\cal I}m\biggl[U^*_{e2} U_{\mu 2} \left(\epsilon^s_{e\mu} |U_{\mu 3}|^2+
\epsilon_{e\tau}^s U_{\mu 3}^*U_{\tau 3}\right) \biggr].
\eea
The last three terms in this expression are CP--violating and would be
the basis for our results.

\section{CP Violation in Vacuum Oscillations}
To measure CP violation, one will need to compare the transition
probability $P_{e\mu}$ evaluated in the previous section to that of
the CP-conjugate process, $P_{\bar e\bar\mu}$. The latter will be
measured in oscillation experiments where antineutrinos are produced
in the source in conjunction with $e^-$ and detected through $\mu^+$
production. It is clear that a CP transformation relates the production 
processes, $\mu^-\to e^-\bar\nu_\alpha\nu_{\alpha^\prime}$ and
$\mu^+\to e^+\nu_\alpha\bar\nu_{\alpha^\prime}$. As concerns the detection
processes, $\bar\nu_\beta u\to\mu^+ d$ and $\nu_\beta d\to\mu^- u$, the
situation is less straightforward. We have $G^d_{\beta\mu}\propto\langle 
p\mu^-|\bar\mu^-\ \bar u\ \nu_\beta\ d|\nu_\beta n\rangle$ and 
$G^d_{\bar\beta\bar\mu}\propto\langle n\mu^+|
\bar\mu^+\ \bar d\ \bar\nu_\beta\ u|\bar\nu_\beta p\rangle$.
The relation is through CP and crossing symmetry, but for a four-fermion 
interaction this is equivalent to a CP transformation.

CP transformation of the Lagrangian takes the elements of the mixing matrix 
and the $\epsilon$-terms into their complex conjugates. It is then 
straightforward to obtain the transition probability for antineutrino 
oscillations. Our interest lies in the CP asymmetry,
\be
A_{\rm CP}={P_-\over P_+},
\label{defacp}
\ee
where
\be
P_\pm=P_{e\mu}\pm P_{\bar e\bar\mu}.
\ee
We quote below the leading contributions for `short' distances,
$x_{31}\ll1$.  In some of the observables, we consider two limiting
cases for $|U_{e3}|$: 
\bea
\label{deflim}
{\mbox{\rm The ``large''}}\ s_{13}\ {\rm limit}&:\ \ \  
&x_{21}/x_{31}\ll|(U_{e3}U_{\mu3})/(U_{e2}U_{\mu2})|, \no\\
{\mbox{\rm The\ small}}\ s_{13}\ {\rm limit}&:\ \ \  
&x_{21}/x_{31}\gg|(U_{e3}U_{\mu3})/(U_{e2}U_{\mu2})|.
\eea

The CP conserving rate $P_+$ is always dominated by the Standard Model.
It is given by
\be
\label{Pplus}
P_+=\cases{8x_{31}^2|U_{e3}U_{\mu3}|^2&large $s_{13}$,\cr
8x_{21}^2|U_{e2}U_{\mu2}|^2&small $s_{13}$.\cr}
\ee

The CP--violating difference between the transition probabilities within
the Standard Model can be obtained from Eq.~(\ref{pemsm}):
\be
\label{Pminsm}
P_-^{\rm SM}=-16
x_{21}x^2_{31}{\cal I}m\bigr(U_{e2}U^*_{\mu 2}U^*_{e3}U_{\mu 3}\bigl).
\ee
As is well known, CP violation within the Standard Model is suppressed
by both the small mixing angle $|U_{e3}|$ and the small mass-squared
difference $\Delta m^2_{12}$. More generally, it is proportional to
the Jarlskog measure of CP violation, $J={\cal I}m\bigr(U_{e2}U^*_{\mu
2}U^*_{e3}U_{\mu 3}\bigl)$.  For short distances
($x_{21},x_{31}\ll1$), the dependence of $P_-^{\rm SM}$ on the distance is
$L^3$. Since it is CP--violating, it should be odd in $L$.  The absence
of a term linear in $L$ comes from the fact that the Standard Model
requires for CP to be violated, that all three mass-squared differences
do not vanish, that is, $P_-\propto\Delta_{21}\Delta_{31}\Delta_{32}$.
In the limit 
$x_{21}/x_{31}\ll|(U_{e3}U_{\mu3})/(U_{e2}U_{\mu2})|$, we obtain the
following Standard Model asymmetry: 
\be
\label{Acpsm}
A_{\rm CP}^{\rm SM}=-2x_{21}{\cal I}m\left({U_{e2}U_{\mu2}^*\over
U_{e3}U_{\mu3}^*}\right).
\ee
In the small $s_{13}$ limit, the standard CP violation is unobservably small.

The CP--violating difference between the transition probabilities that
arises from the New Physics interactions can be obtained from 
Eq.~(\ref{pemnp}):
\be
\label{Pminnp}
P_-^{\rm NP}=\cases{-8x_{31}{\cal I}m[U_{e3}^*U_{\mu 3}
(\epsilon^{d*}_{\mu e}+\epsilon^s_{e\mu})]&large $s_{13}$,\cr
-8 x_{21}{\cal I}m[U_{e2}^* U_{\mu 2}
(\epsilon^{d*}_{\mu e}+\epsilon_{e\mu}^s)]&small $s_{13}$.\cr} 
\ee
We learn that CP violation beyond the weak interactions requires only
that either $|U_{e3}|$ or $\Delta m^2_{21}$ be different from zero,
but not necessarily both. Also the dependence on the distance is
different: for short distances, $P_-^{\rm NP}\propto L$. From
Eqs.~(\ref{Pplus}) and~(\ref{Pminnp}) we obtain the following new
physics contribution to the CP asymmetry:
\be
\label{Acpnp}
A_{\rm CP}^{\rm NP}=\cases{-{1\over x_{31}}{\cal I}m\left(
{\epsilon^{d*}_{\mu e}+
\epsilon^s_{e\mu}\over U_{e3}U_{\mu3}^*}\right)&large $s_{13}$,\cr
-{1\over x_{21}}{\cal I}m\left({\epsilon^{d*}_{\mu e}+
\epsilon^s_{e\mu}\over U_{e2}U_{\mu2}^*}\right)&small $s_{13}$.\cr}
\ee
The apparent divergence of $A_{\rm CP}^{\rm NP}$ for small $L$ is only
due to the approximations that we used. Specifically, there is an 
${\cal O}(|\epsilon|^2)$ contribution to $P_+$ that is constant
in $L$ \cite{Yuval}, namely $P_+={\cal O}(|\epsilon|^2)$ for $L\to0$.
In contrast, $P_-=0$ in the $L\to0$ limit to all orders in $|\epsilon|$.

Equations (\ref{Pminnp}) and (\ref{Acpnp}) lead to several interesting
conclusions:

(i) It is possible that, in CP--violating observables, the New Physics
contributions compete with or even dominate over the Standard Model ones 
in spite of the superweakness of the interactions ($|\epsilon|\ll1$). Given 
that for the proposed experiments $x_{31}\lsim1$, it is sufficient that
\be
\label{largenp}
{\rm max}\left(|\epsilon^s_{e\mu}|,|\epsilon^d_{\mu e}|\right)\geq
{\rm min}\left(|U_{e3}|,x_{21}\right),
\ee
for the new contribution to the CP--violating difference $P_-$
to be larger than the standard one.

(ii) The different distance dependence of 
$P_-^{\rm NP}$ and $P_-^{\rm SM}$
will allow, in principle, an unambiguous distinction to be made 
between New Physics contributions of the type described here and the
contribution from lepton mixing.

(iii) The $1/L$ dependence of $A_{\rm CP}^{\rm NP}$ suggests that the optimal
baseline to observe CP violation from New Physics is shorter than the one
optimized for the Standard Model.

We carried out a numerical calculation of the probabilities $P_\pm$ and
asymmetry $A_{CP}$ as a function of the distance between the source and the
detector. We use $E_\nu=20$ GeV, which is the range of neutrino energy
expected in neutrino factories. For the neutrino parameters, we take
$\Delta m^2_{31}=3\times10^{-3}$ eV$^2$ and $\tan^2\theta_{23}=1$, consistent
with the atmospheric neutrino measurements \cite{3nus}, and 
$\Delta m^2_{21}=10^{-4}$ eV$^2$ and $\tan^2\theta_{12}=1$, consistent at 
present with the LMA solution of the solar neutrino problem \cite{3nus,nu2000}. As concerns the third mixing angle and CP--violating phase in the lepton mixing 
matrix, we consider two cases. First, we take $s_{13}=0.2$, close to the upper 
bound from CHOOZ \cite{chooz,pave,3nus}, and $\delta=\pi/2$. This set of 
parameters is the one that maximizes the standard CP asymmetry. Second, we take 
$s_{13}=0$, in which case there is no standard CP violation in the lepton 
mixing. As concerns the effects of New Physics, we demonstrate them by taking 
only $|\epsilon^s_{e\mu}|\neq 0$. With our first set of mixing parameters 
(maximal standard CP violation),  we take $|\epsilon^s_{e\mu}|=10^{-3}$ and 
$\delta_\epsilon=0$. With our second set of mixing parameters (zero standard CP 
violation),  we take $|\epsilon^s_{e\mu}|=10^{-4}$ and $\delta_\epsilon=\pi/2$.
Our choice of CP--violating phases can be easily understood on the basis of
Eq.~(\ref{Acpnp}): in the large $s_{13}$ limit, the CP asymmetry depends
on $\arg[\epsilon^s_{e\mu}/(U_{e3}U_{\mu3}^*)]=\delta_\epsilon-\delta$,
while in the small $s_{13}$ limit it depends on 
$\arg[\epsilon^s_{e\mu}/(U_{e2}U_{\mu2}^*)]=\delta_\epsilon$.
We use the full expression for the transition probabilities that is 
presented in  Appendix A. Consequently, the only approximation that 
we make is that we omit  effects of  ${\cal O}(|\epsilon|^2)$. 

\begin{figure}
\centerline{\mbox{\epsfig{file=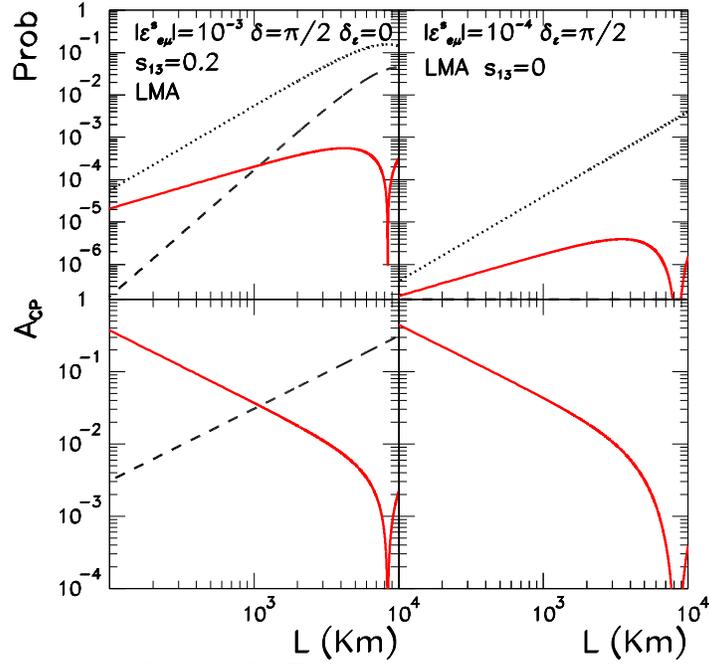,width=0.6\textwidth}}}
\caption{
Transition probabilities and CP asymmetries in vacuum as a function of the 
distance. In the upper panels the curves correspond to $P_+^{\rm SM}$
(dotted), $P_-^{\rm SM}$ (dashed) and $P_-^{\rm NP}$ (solid). In the lower 
panels the curves correspond to $A_{\rm CP}^{\rm NP}$ (solid) and 
$A_{\rm CP}^{\rm SM}$ 
(dashed). In the left panels, $s_{13}=0.2$, $\delta=\pi/2$, 
$|\epsilon^s_{e\mu}|=10^{-3}$ and $\delta_\epsilon=0$. In the right panels, 
$s_{13}=0$, $|\epsilon^s_{e\mu}|=10^{-4}$ and $\delta_\epsilon=\pi/2$.
In all curves $E_\nu=20$ GeV, $\Delta m^2_{13}=3\times 10^{-3}$ 
eV$^2$, $\tan^2\theta_{23}=1$, $\Delta m^2_{21}=10^{-4}$ 
eV$^2$ and $\tan^2\theta_{12}=1$.}
\label{cpvac}
\end{figure}

The results of this calculation are presented in Fig.~\ref{cpvac}. The left 
panels correspond to the first case (maximal standard CP violation) and the 
right ones to the second (zero standard CP violation). For each case we 
present, as a function of the distance between the source and the detector, 
$P_+$ (dotted line), $P_-^{\rm SM}$ and $A_{CP}^{\rm SM}$ (dashed lines in,
respectively, upper and lower panels), and $P_-^{\rm NP}$ and 
$A_{CP}^{\rm NP}$ (solid lines in, respectively, upper and lower panels).

We learn a few interesting facts:

(i) The New Physics contribution to CP violation can dominate over
even the maximal standard  CP violation for values of
$|\epsilon|$ as small as $10^{-4}$. This is particularly valid for
distances shorter than 1000 km.

(ii) The approximations that lead to Eqs.~(\ref{Pminnp}) and
(\ref{Acpnp}) are good for $L\lsim5000$ km.

(iii) As anticipated from our approximate expressions, for short enough
distances, $P_-^{\rm NP}$ grows linearly with 
distances and $A_{CP}^{\rm NP}$
is strongly enhanced at short distances.

(iv) In the large $s_{13}$ case, the new CP violation is sensitive mainly
to the phase difference $\delta-\delta_\epsilon$ and is almost independent
of the solar neutrino parameters.

(v) In the very small $s_{13}$ limit, the new CP violation is
proportional to $\sin\delta_\epsilon$. The rate $P_-^{\rm NP}$ 
is suppressed by the solar neutrino mass difference and mixing angle.

\section{The Transition Probability in Matter}
Since long--baseline experiments involve the propagation of neutrinos in
the matter of Earth, it is important to understand matter effects on
our results. For our purposes, it is sufficient to study the case of
constant matter density. Then the matter contribution to the effective
$\nu_e$ mass, $A=\sqrt{2}G_F N_e$, is constant. 

One obtains the transition probability in matter by replacing the
mass-squared differences $\Delta_{ij}$ and mixing angles $U_{\alpha
i}$ with their effective values in matter, $\Delta_{ij}^m$ and
$U_{\alpha i}^m$. The full expression for $P_{e\mu}$ in matter can
then be written in terms of $x_{ij}^m$ and $U_{\alpha i}^m$ by a
straightforward modification of the vacuum probability given in
Appendix A. To understand the matter effects it is, however, more
useful to take into account the smallness of $|U_{e3}|$ and $x_{12}$.
We will expand the transition probability in these parameters to
second order for $P_{e\mu}^{\rm SM}$ and to first order for
$P_{e\mu}^{\rm NP}$.

For the Standard Model case, we obtain:
\bea
\label{psmmat}
P_{e\mu}^{\rm SM}&=& 4 \left(\frac{\Delta_{21}}{A}\right)^2
\sin^2\left(\frac{AL}{2}\right)|U_{e2}U_{\mu 2}|^2+
4 \left(\frac{\Delta_{31}}{B}\right)^2\sin^2\left(\frac{BL}{2}\right)
|U_{e3}U_{\mu 3}|^2 \no \\
&& +8\left(\frac{\Delta_{21}}{A}\right)\left(\frac{\Delta_{31}}{B}\right)
\sin\left(\frac{AL}{2}\right)\sin\left(\frac{BL}{2}\right) \\
&&\ \bigg\{\cos x_{31}{\cal R}e \bigl[U_{e3}^* U_{\mu3} U_{e2} U_{\mu 2}^* \bigr] 
-\sin x_{31}{\cal I}m\bigl[U_{e3}^*U_{\mu3}U_{e2}U_{\mu 2}^*\bigr]\bigg\}, \no
\eea
where
\be
\label{defB}
B=\Delta_{31}-A.
\ee
Again, the first term is the full result for two generations, and the
second is the full result for the case of $\Delta_{21}=0$. The last
term violates CP. In the limit $A=0$, Eq.~(\ref{pemsm}) is
reproduced. Note that our definition of $B$ is such that $B$ changes
sign according to whether $\Delta_{31}$ is larger or smaller than
$A$. This is different from the usual convention where
$B=|\Delta_{31}-A|$. The Standard Model results are an even function
of $B$ and either definition can be used. But for the New Physics
results given below, the choice of convention is important.

For the New Physics contribution we find:
\bea
\label{pnpmat}
P_{e\mu}^{\rm NP}&=& 4\left(\frac{\Delta_{21}}{A}\right)
\sin^2\left(\frac{AL}{2}\right){\cal R}e \biggl[U^*_{e2}U_{\mu 2}
\left(\epsilon_{\mu e}^{d *}-\epsilon_{e\mu}^{s}(1-2|U_{\mu 3}|^2)
+2\epsilon^s_{e\tau}U^*_{\mu 3} U_{\tau 3}\right)\biggr]\no  \\ 
&-&4\left(\frac{\Delta_{31}}{B}\right)\sin^2\left(\frac{BL}{2}\right)
{\cal R}e\biggl[U^*_{e3}U_{\mu 3}\left(\epsilon_{\mu e}^{d *}+
\epsilon_{e\mu}^{s}(1-2|U_{\mu 3}|^2)-2\epsilon^s_{e\tau}
U^*_{\mu 3} U_{\tau 3}\right)\biggr] \no \\  
&-&2\left(\frac{\Delta_{21}}{A}\right) \sin(AL){\cal I}m
\biggl[U_{e2}^* U_{\mu 2}(\epsilon^{*d}_{\mu e}+\epsilon^s_{e\mu})\biggr]\no\\
&-&2\left(\frac{\Delta_{31}}{B}\right) \sin(BL){\cal I}m
\biggl[U_{e3}^* U_{\mu 3}(\epsilon^{*d}_{\mu e}+\epsilon^s_{e\mu})\biggr]\no \\
&-&8 \sin\left(\frac{AL}{2}\right)\sin\left(\frac{BL}{2}\right)\cos x_{31}\no\\ 
&& \biggl\{\left(\frac{\Delta_{31}}{B}\right)
{\cal R}e\biggl[U_{e3}^*U_{\mu 3}\left(\epsilon^s_{e\mu}(1-|U_{\mu 3}|^2)
-\epsilon^s_{e\tau}U^*_{\mu 3} U_{\tau 3}\right)\biggr] \no \\
&&-\left(\frac{\Delta_{21}}{A}\right)
{\cal R}e\biggl[U_{e2}^*U_{\mu 2}\left(\epsilon^{s}_{e\mu}|U_{\mu 3}|^2+
\epsilon^s_{e\tau}U^*_{\mu 3} U_{\tau 3}\right)\biggr] \biggr\}\no \\
&+&8 \sin\left(\frac{AL}{2}\right)\sin\left(\frac{BL}{2}\right)\sin x_{31}\no\\
&&\bigg\{\left(\frac{\Delta_{31}}{B}\right)
{\cal I}m\biggl[U_{e3}^*U_{\mu 3}\left(\epsilon^s_{e\mu}(1-|U_{\mu 3}|^2)
-\epsilon^s_{e\tau}U^*_{\mu 3} U_{\tau 3}\right)\biggr] \no \\
&&-\left(\frac{\Delta_{21}}{A}\right)
{\cal I}m\biggl[U_{e2}^*U_{\mu 2}\left(\epsilon^{s}_{e\mu}|U_{\mu 3}|^2+
\epsilon^s_{e\tau}U^*_{\mu 3} U_{\tau 3}\right)\biggr]\biggr\}.
\eea 

Unlike the case of vacuum oscillation, $P_-$ will get contributions
from both CP--violating terms (proportional to the imaginary parts of
various combinations of parameters) as CP conserving terms
(proportional to the real parts).

Note that, in addition to the effects of new neutrino interactions in the 
source and in the detector, there could be other, independent effects due to 
new neutrino interactions with matter during their propagation 
\cite{fcni1,fcni2,fcni3}. Such effects have been studied in the
context of solar and atmospheric neutrinos (see {\it e.g.} 
Refs.~\cite{fcnisun1,fcnisun2,fcniatm}) but we neglect them here. 
 
\section{CP Violation in Matter Oscillations}
Since matter in Earth is not CP symmetric, there will be contributions
to $A_{CP}$ even in the case when there is no CP violation. It is our
purpose in this section to evaluate these contributions and, in
particular, the fake asymmetry that is related to the real part of
$\epsilon$.  We denote the matter--related contribution to $P_-$ by
$P_-^m\equiv P_-(A)-P_-(A=0)$. Since the leading contributions to $P_+$
are the same as in the vacuum case [Eq.~(\ref{Pplus})], we can similarly
define the matter--related contribution to $A_{\rm CP}$: $A_{\rm CP}^m\equiv 
P_-^m/P_+$. Note that in the evaluation of
$P_{\bar e\bar\mu}$ from the expressions that we found for $P_{e\mu}$
we need not only to replace $U_{\alpha i}$ and
$\epsilon^{s,d}_{\alpha\beta}$ with their complex conjugates, but also
$A$ with $-A$.

For the Standard Model, we obtain from Eq.~(\ref{psmmat}), in the
small $x_{31}$ and large $s_{13}$ limits,
\be
\label{pmsmmat}
(P_-^m)^{\rm SM}={16\over3}x_{31}^4\left({A\over\Delta_{31}}\right)
|U_{e3}U_{\mu3}|^2.
\ee
In the small $s_{13}$ limit ($x_{21}/x_{31}\gg
|(U_{e3}U_{\mu3})/(U_{e2}U_{\mu2})|$) the Standard Model effect is 
unobservably small, and we do not consider it here.
Taking into account that [see Eq.~(\ref{Pplus})]
$P_+\approx8x_{31}^2|U_{e3}U_{\mu3}|^2$, we get
\be
\label{asmmat}
(A_{CP}^m)^{\rm SM}={2\over3}x_{31}^2\left({A\over\Delta_{31}}\right).
\ee

For the New Physics contribution, we obtain from Eq.~(\ref{pnpmat}), 
in the small $x_{31}$ limit,
\be
\label{pmnpmat}
(P_-^m)^{\rm NP}=\cases{
8x_{31}^2{A\over\Delta_{31}}{\cal R}e[U_{e3}^*U_{\mu3}
(\epsilon^{d*}_{\mu e}-\epsilon^s_{e\mu})]&large $s_{13}$,\cr
8x_{21}^2{A\over\Delta_{21}}{\cal R}e[U_{e2}^*U_{\mu2}
(\epsilon^{d*}_{\mu e}-\epsilon^s_{e\mu})]&small $s_{13}$,\cr}
\ee
and
\be
\label{anpmat}
(A_{CP}^m)^{\rm NP}=\cases{
{A\over\Delta_{31}}{\cal R}e\left({\epsilon^{d*}_{\mu e}-\epsilon^s_{e\mu}
\over U_{e3}U_{\mu3}^*}\right)&large $s_{13}$,\cr
{A\over\Delta_{21}}{\cal R}e\left({\epsilon^{d*}_{\mu e}-\epsilon^s_{e\mu}
\over U_{e2}U_{\mu2}^*}\right)&small $s_{13}$.\cr}
\ee

We would like to make a few comments regarding our results here:

(i) Each of the four contributions has a different dependence on the 
distance. In the short distance limit, we have
\be
(P_-^m)^{\rm SM}\propto L^4,\ \ \ P_-^{\rm SM}\propto L^3,\ \ \ 
(P_-^m)^{\rm NP}\propto L^2,\ \ \ P_-^{\rm NP}\propto L,
\ee
and, equivalently,
\be
(A_{\rm CP}^m)^{\rm SM}\propto L^2,\ \ \ A_{\rm CP}^{\rm SM}\propto L,\ \ \ 
(A_{\rm CP}^m)^{\rm NP}\propto L^0,\ \ \ A_{\rm CP}^{\rm NP}\propto 1/L.
\ee
One can then distinguish between the various contributions, at least
in principle.

(ii) If the phases of the $\epsilon$'s are of order 1, then the genuine
CP asymmetry will be larger (at short distances) than the fake one.

(iii) It is interesting to note that the search for CP violation in
neutrino oscillations will allow us to constrain both 
${\cal R}e(\epsilon)$ and ${\cal I}m(\epsilon)$.

We carried out a numerical calculation of the probabilities $P_\pm^m$
and asymmetry $A_{\rm CP}^m$ as a function of the distance between the
source and the detector. We use again $E_\nu=20$ GeV, $\Delta
m^2_{31}=3\times10^{-3}$ eV$^2$, $\tan^2\theta_{23}=1$, $\Delta
m^2_{21}=10^{-4}$ eV$^2$, $\tan^2\theta_{12}=1$ and $s_{13}=0.2$ or
$0$. For the New Physics parameters, we take
$|\epsilon^s_{e\mu}|=10^{-3}$. To isolate the matter effects we now,
however, switch off all genuine CP violation, that is, we take
$\delta=\delta_\epsilon= 0$ in both cases.

\begin{figure}
\centerline{\mbox{\epsfig{file=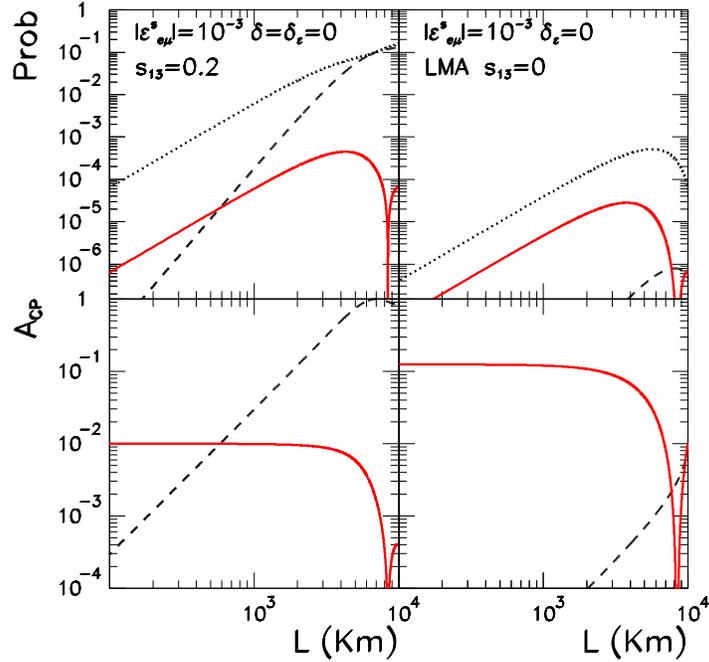,width=0.6\textwidth}}}
\caption{Transition probabilities and fake CP asymmetries in matter as a function 
of the distance. All CP--violating phases are set to zero. In the upper panels 
the curves correspond to $P_+^{\rm SM}$ (dotted), $(P_-^m)^{\rm SM}$ (dashed) 
and $(P_-^m)^{\rm NP}$ (solid). In the lower panels the curves correspond to 
$(A_{\rm CP}^m)^{\rm NP}$ (solid) and $(A_{\rm CP}^m)^{\rm SM}$ (dashed). 
In the left 
panels $s_{13}=0.2$, and in the right panels $s_{13}=0$. In all curves 
$E_\nu=20$ GeV, $\Delta m^2_{31}=3\times 10^{-3}$ eV$^2$, $\tan^2\theta_{23}=1$,
$\Delta m^2_{21}=10^{-4}$ eV$^2$, $\tan^2\theta_{12}=1$, 
$\delta=\delta_\epsilon=0$ and $|\epsilon^s_{e\mu}|=10^{-3}$.}
\label{cpmat}
\end{figure}

The results of this calculation are presented in Fig.~\ref{cpmat}. The
left panels correspond to the first case (large $s_{13}$) and the
right ones to the second (vanishing $s_{13}$). For each case we
present, as a function of the distance between the source and the
detector, $P_+$ (dotted line), $(P_-^m)^{\rm SM}$ and $(A_{CP}^m)^{\rm
SM}$ (dashed lines in, respectively, upper and lower panels), and
$(P_-^m)^{\rm NP}$ and $(A_{CP}^m)^{\rm NP}$ (solid lines in,
respectively, upper and lower panels).

We learn a few interesting facts:

(i) The New Physics contribution to the fake CP violation can dominate
over the standard  contribution for values of $|\epsilon|$ as
small as $10^{-4}$. This is particularly valid for distances shorter
than 500 km.

(ii) As anticipated from our approximate expressions, for short enough
distances $(P_-^m)^{\rm NP}$ grows quadratically with distances and 
$(A_{CP}^m)^{\rm NP}$ is independent of the distance.

(iii) Both the standard and the new contribution to $P_-^m$ are
suppressed by a small $s_{13}$. The $s_{13}$ suppression is however
stronger for $P_+$ than it is for $(P_-^m)^{\rm NP}$. Consequently,
the New Physics contribution to $(A_{CP}^m)^{\rm NP}$ becomes very
large for vanishing $s_{13}$.

\begin{figure}
\centerline{\mbox{\epsfig{file=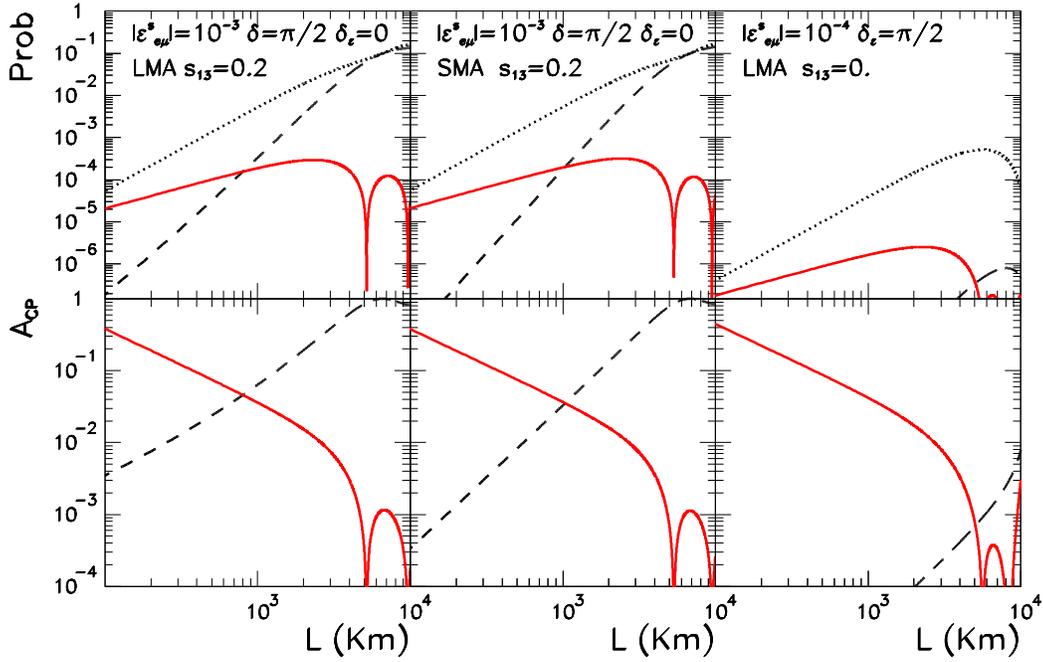,width=0.9\textwidth}}}
\caption{
Transition probabilities and fake CP asymmetries in matter as a function of 
the distance. In the upper panels the curves correspond to $P_+^{\rm SM}$ 
(dotted), $P_-^{\rm SM}$ (dashed) and $P_-^{\rm NP}$ (solid). In the lower 
panels the curves correspond to $A_{\rm CP}^{\rm NP}$ (solid) and 
$A_{\rm CP}^{\rm SM}$ (dashed). In all curves $E_\nu=20$ GeV, $\Delta m^2_{31}=
3\times 10^{-3}$ eV$^2$ and $\tan^2\theta_{23}=1$. In the left panels 
$\Delta m^2_{21}=10^{-4}$ eV$^2$, $\tan\theta_{12}=1$, $s_{13}=0.2$,
$\delta=\pi/2$, $|\epsilon^s_{e\mu}|=10^{-3}$ and $\delta_\epsilon=0$. 
In the middle panels $\Delta m^2_{21}=5.2\times10^{-6}$ eV$^2$,
$\tan^2\theta_{12}=7.5\times10^{-4}$, $s_{13}=0.2$, $\delta=\pi/2$,
$|\epsilon^s_{e\mu}|=10^{-3}$ and $\delta_\epsilon=0$. In the right panels
$\Delta m^2_{21}=10^{-4}$ eV$^2$, $\tan\theta_{12}=1$, $s_{13}=0$,
$|\epsilon^s_{e\mu}|=10^{-4}$ and $\delta_\epsilon=\pi/2$.}
\label{cptot}
\end{figure}

In reality, the measured $P_-$ and $A_{CP}$ will be affected by both
genuine CP--violating contributions and matter-induced
contributions. This situation is illustrated in Fig.~\ref{cptot}. We
present $P_+$ (dotted curve), $P_-^{\rm SM}$ and $A_{CP}^{\rm SM}$
(dashed curves in, respectively, upper and lower panels), and
$P_-^{\rm NP}$ and $A_{CP}^{\rm NP}$ (solid curves in, respectively,
upper and lower panels), as a function of the distance.  For the
neutrino parameters, we always take $\Delta m^2_{31}=3\times10^{-3}$
eV$^2$ and $\tan\theta_{23}=1$, consistent with the atmospheric
neutrino data.  For the other parameters, we take three cases: (a)
Left panel: we take the LMA parameters ($\Delta m^2_{21}=10^{-4}$
eV$^2$ and $\tan\theta_{12}=1$), `large' $s_{13}=0.2$ and maximal
phase $\delta=\pi/2$. This choice of parameters gives maximal standard
CP violation. For the New Physics parameters we take
$|\epsilon^s_{e\mu}|=10^{-3}$ and $\delta_\epsilon=0$. (The reason for
the choice of phase is that the dominant contributions depend on
$\delta-\delta_\epsilon$.) (b) Middle panel: we take the SMA
parameters ($\Delta m^2_{21}=5.2\times10^{-6}$ eV$^2$ \cite{3nus,nu2000},
$\tan^2\theta_{12}=7.5\times10^{-4}$), $s_{13}=0.2$, $\delta=\pi/2$,
$|\epsilon^s_{e\mu}|=10^{-3}$ and $\delta_\epsilon=0$. Here the
standard CP violation is unobservably small, but the standard matter
effects are still large. (c) Right panel: we take the LMA parameters
and $s_{13}=0$.  With a vanishing $s_{13}$, the total transition
probability is highly suppressed as is the standard matter effect, and
standard CP violation vanishes. For the New Physics parameters we take
$|\epsilon^s_{e\mu}|=10^{-4}$ and $\delta_\epsilon=\pi/2$. We take a
smaller $|\epsilon^s_{e\mu}|$ so that our approximation will not
break down.

\begin{figure}
\centerline{\mbox{\epsfig{file=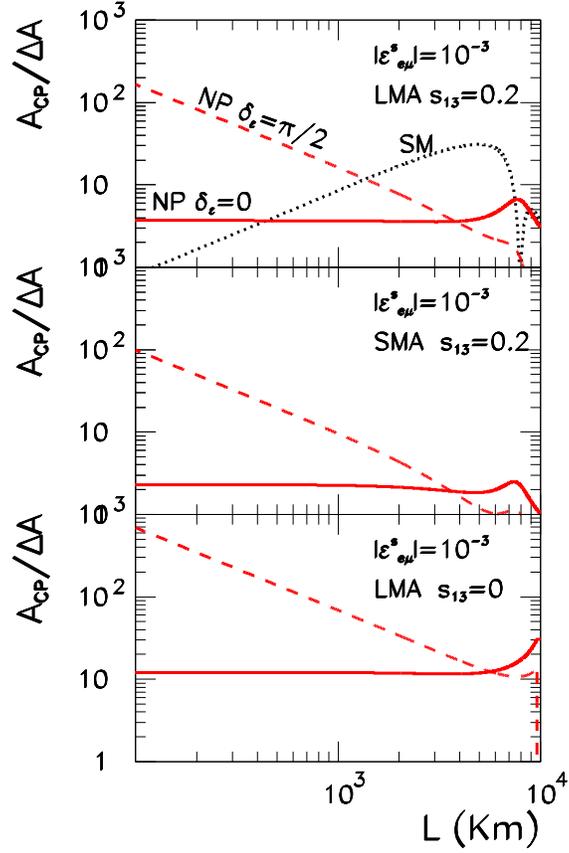,width=0.5\textwidth}}}
\caption{The signal-to-noise ratio, $\overline{A_{\rm CP}^{\rm NP}}/\Delta A$, 
as a function of the distance $L$. We considered the following parameters
for the experiment: $E_\mu=50$ GeV, $10^{21}$ $\mu^-$ decays and a 40 kt 
detector, and the following neutrino parameters: $\delta=0$,
$\Delta m^2_{31}=3\times10^{-3}$ eV$^2$, $\tan\theta_{23}=1$.
In the upper and lower (middle) panels we use the LMA (SMA) parameters.
In the upper two (lower) panels we use $s_{13}=0.2(0)$.
For the New Physics we take $|\epsilon^s_{e\mu}|=10^{-3}$ and
$\delta_\epsilon=0$ or $\pi/2$. In the upper panel, the dotted curve gives
the SM matter-subtracted asymmetry $A_{\rm CP}^{\rm SM}(\delta=\pi/2)-
A_{\rm CP}^{\rm SM}(\delta=0)$.}
\label{sensi}
\end{figure}

We would like to emphasize the following points:

(i) Similar three cases will be the basis, in the next section, 
for our 
analysis of the sensitivity of CP--violating observables measured in neutrino 
factories to New Physics effects (see Fig.~\ref{sensi}).

(ii) With large $s_{13}$, the dependence of the New Physics effects (and of
the standard matter-induced effects) on the solar neutrino parameters is
very weak. 

(iii) A small or even vanishing $s_{13}$ will suppress all the rates
and will introduce a strong dependence on the solar neutrino
parameters. The New Physics contributions to $A_{\rm CP}$ will be,
however, only little affected because both the standard CP conserving
rate and the New Physics CP--violating rate are suppressed in the same
way.

(iv) With large $s_{13}$, the New Physics CP--violating effects are dominated
by the combination $\delta-\delta_\epsilon$. With small (but not vanishing)
$s_{13}$, the dependence is on both $\delta-\delta_\epsilon$ and 
$\delta_\epsilon$. 

(v) For distances shorter than 800 km, the effects of
$|\epsilon|\gsim10^{-3}$ are always dominant. For distances shorter
than 300 km, the New Physics dominates even for
$|\epsilon|\sim10^{-4}$.

\section{Long--Baseline Experiments}
We would like to quantify the sensitivity of a neutrino factory to the
CP--violating effects from new neutrino interactions. For this purpose,
we consider the measurement of the following integrated asymmetry
\cite{dghr}:
\begin{equation}
\overline{A_{CP}}=\frac
{\left.{N[\mu^-]}/{N_0[e^-]}\right|_+-\left.{N[\mu^+]}/{N_0[e^+]}\right|_-}
{\left.{N[\mu^-]}/{N_0[e^-]}\right|_++\left.{N[\mu^+]}/{N_0[e^+]}\right|_-}.
\label{acpin}
\end{equation}
Here  $N[\mu^-]/N_0[e^-]|_+$ refers to an oscillation experiment that has
$\mu^+$ decay as its production process:  $N[\mu^-]$ is the measured number of 
wrong--sign muons while $N_0[e^-]$ is the expected number of $\nu_e$ CC 
interaction events (in the absence of oscillations). Similarly, 
$N[\mu^+]/N_0[e^+]|_-$ refers to an oscillation experiment that has $\mu^-$ 
decay as its production process:  $N[\mu^+]$ is the measured number of 
wrong--sign muons while $N_0[e^+]$ is the expected number of 
$\overline{\nu_e}$ CC 
interaction events (again, in the absence of oscillations). 
The measured number of wrong--sign muon events can be expressed as follows:
\begin{equation}
N[\mu^-]|_+=\frac{N_\mu N_T }{\pi m_\mu^2} 
\frac{E_\mu}{L^2}\int dE_\nu f_\nu(E_\nu) \sigma_{CC}(E_\nu)P_{e\mu}(E_\nu),
\label{wsm}
\end{equation}
where $N_T$ is the number of protons in the target detector, $N_\mu$
is the number of useful muon decays, $E_\mu$ is the muon energy and
$m_\mu$ is the muon mass. The function $f_\nu(E_\nu)$ is the energy
distribution of the produced neutrinos. We assume that the muons are
not polarized, in which case $f_\nu(E_\nu)=12 x^2 (1-x)$ with
$x=E_\nu/E_\mu$. Finally, $\sigma_{CC}(E_\nu)$ is the neutrino--nucleon
interaction cross section which, in the interesting range of energies,
can be taken to be proportional to the neutrino energy:
$\sigma_{CC}=\sigma_0 E_\nu$ with $\sigma_0=0.67 \times 10^{-38}$
cm$^2$/GeV for neutrinos and $\sigma_0=0.34 \times 10^{-38}$
cm$^2$/GeV for antineutrinos.  The expression for $N[e^-]|_+$ is
obtained by an integral similar to Eq.~(\ref{wsm}), except that
$P_{e\mu}$ is replaced by $1$.

We define $\overline{A_{\rm CP}^{\rm NP}}$ as the contribution from new
physics (that is, $\epsilon$-dependent) to the integrated CP
asymmetry. We take into account both genuine CP--violating and
matter-induced contributions. (In the limit of a real lepton mixing
matrix, that is, no standard CP violation, the first contributions are
proportional to ${\cal I}m(\epsilon)$ and the latter to ${\cal
R}e(\epsilon)$.) We define $\Delta A$ to be the statistical error on
$\overline{A_{\rm CP}}$. In order to quantify the significance of the
signal due to New Physics, we compute the ratio $\overline{A_{\rm CP}^{\rm
NP}}/\Delta A$.

The statistical error, $\Delta A$, scales with distance and energy as follows:
\begin{equation}
\Delta A \simeq \frac{1}{\sqrt{N[\mu^+]|_-+N[\mu^-]|_+}}\propto
\frac{1}{\sqrt{P_+^{\rm SM} N_{CC}}}\propto \frac{1}{\sqrt{E_\nu}}. 
\label{delA} 
\end{equation}
To find this scaling, we took into account that the number of CC
interactions scales as $N_{CC}\propto E_\nu^3/L^2$ while, for $L\lsim 3000$ km,
$P_+^{SM}\propto L^2/E^2_\nu$. Consequently, the dependence of $\Delta A$
on the distance is very weak. Given our results for $A_{\rm CP}^{\rm NP}$, we 
obtain the following scaling with distance of the signal-to-noise ratio:
\begin{equation}
\overline{A_{\rm CP}^{\rm NP}}/\Delta A\propto\cases{
1/L&genuine CP--violating effects,\cr
{\rm const}(L)&matter induced effects.\cr}  
\end{equation}

This behavior is illustrated in Fig.~\ref{sensi} where we display the
signal-to-noise ratio, $\overline{A_{\rm CP}^{\rm NP}}/{\Delta A}$, as a
function of the distance. For simplicity, we consider only the effect
of $\epsilon^s_{e\mu}$. The standard CP violation is presented only in
the upper panel, where it corresponds to maximal $A_{CP}^{\rm SM}$
(LMA parameters: $\Delta m^2_{21}=10^{-4}$ eV$^2$ and
$\tan\theta_{12}=1$, large $s_{13}$ and $\delta=\pi/2$), while the
middle panel has unobservably small $A_{\rm CP}^{\rm SM}$ (SMA parameters:
$\Delta m^2_{21}=5.2\times 10^{-6}$ eV$^2$ and
$\tan^2\theta_{12}=7.5\times10^{-4}$), and the lower panel has zero
$A_{\rm CP}^{\rm SM}$ ($s_{13}=0$). As concerns the new CP violation, the
dashed line corresponds to the case with maximal CP--violating phase
($\delta_\epsilon=\frac{\pi}{2}$) and the solid line corresponds to
purely matter-induced asymmetry ($\delta_\epsilon=0$).  In our
calculations we have assumed a total of $10^{21}$ useful $\mu^-$
decays with energy $E_\mu=50$ GeV and a 40 kt detector.

It is clear from the figure that the maximal sensitivity to new, 
CP--violating contributions to the production or detection processes will
be achieved with shorter distances, while the sensitivity to CP
conserving contributions through matter induced effects is almost
independent of distance.

A truly short baseline experiment can potentially probe the 
${\cal O}(|\epsilon|^2)$ CP conserving effects.  But in this case, due to
the small signal, systematic errors will dominate over the statistical
ones discussed above. It is unlikely that $|\epsilon|$ smaller than
${\cal O}(10^{-3})$ can be signalled in such a measurement.

\begin{figure}
\centerline{\mbox{\epsfig{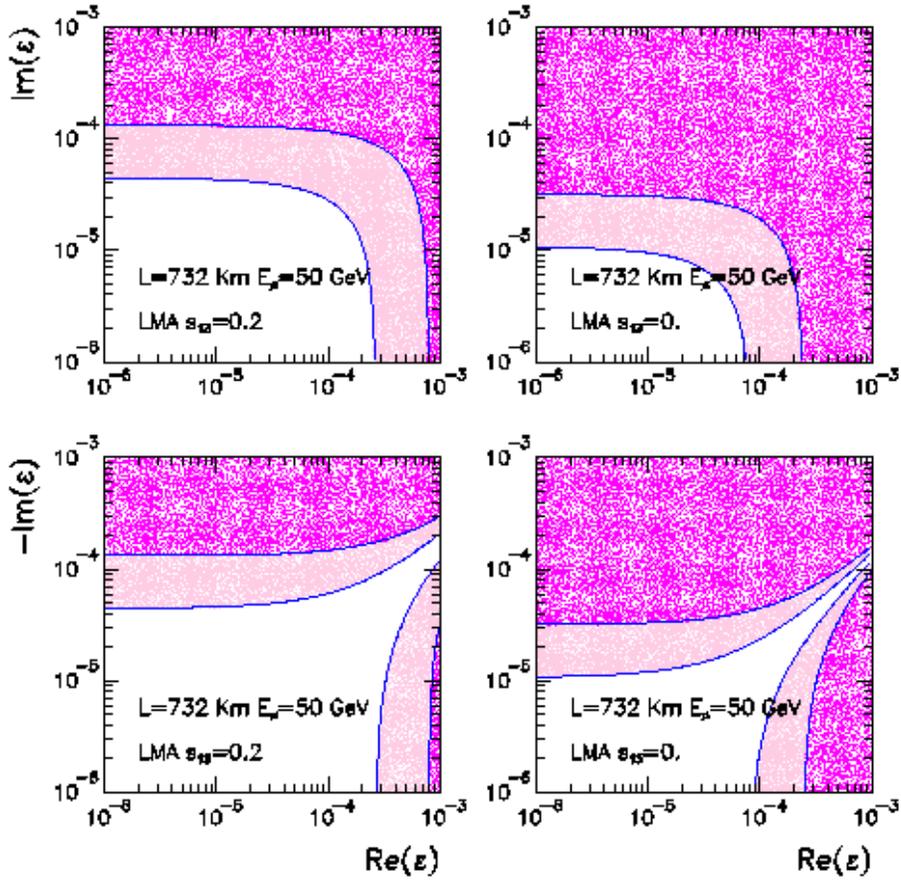}}}
\caption{Regions in the plane of [${\cal R}e(\epsilon^s_{e\mu}),
{\cal I}m(\epsilon^s_{e\mu})$] that give 
$\overline{A_{\rm CP}^{\rm NP}}/\Delta A=3$
(darker shadow) and 1 (light shadow). For the experiment, we take 
$L=732$ km, $E_\mu=50$ GeV, $10^{21}$ $\mu^-$ decays and a 40 kt detector.
For the neutrino parameters, we take $\delta=0$, 
$\Delta m^2_{31}=3\times 10^{-3}$ eV$^2$, $\tan^2\theta_{23}=1$, 
$\Delta m^2_{21}=10^{-4}$ eV$^2$ and $\tan^2\theta_{12}=1$.
In the left (right) panels we have $s_{13}=0.2 (0)$.}
\label{cont732}
\end{figure}

We next investigate the sensitivity to the size of the New Physics
interaction that can be achieved by the measurement of the integrated
CP asymmetry. In Fig.~\ref{cont732}, we show the regions in the
[${\cal R}e(\epsilon^s_{e\mu}), {\cal I}m(\epsilon^s_{e\mu})$] plane
that will lead to $\overline{A_{\rm CP}^{\rm NP}}/{\Delta A}=3$ (
darker--shadow region) and $\overline{A_{\rm CP}^{\rm NP}}/{\Delta A}=1$
(lighter--shadow regions) at $L=732$ km, the shorter baseline discussed
for an oscillation experiment at a neutrino factory. We have assumed a
total of $10^{21}$ useful $\mu^-$ decays with energy $E_\mu=50$ GeV
and a 40 kt detector. In all panels we have $\delta=0$ (no standard CP
violation), $\Delta m^2_{31}=3\times10^{-3}$ eV$^2$ and
$\tan\theta_{23}=1$, and the LMA parameters, $\Delta m^2_{21}=10^{-4}$
eV$^2$ and $\tan\theta_{12}=1$. In the left panels we have
$s_{13}=0.2$ and in the right ones $s_{13}=0$. In the upper panels
${\cal I}m(\epsilon^s_{e\mu})>0$, which, for our choice of parameters,
results in a constructive interference between the matter-induced and
CP--violating effects, while in the lower panels ${\cal
I}m(\epsilon^s_{e\mu})<0$, which results in a destructive
interference.

\begin{figure}
\centerline{\mbox{\epsfig{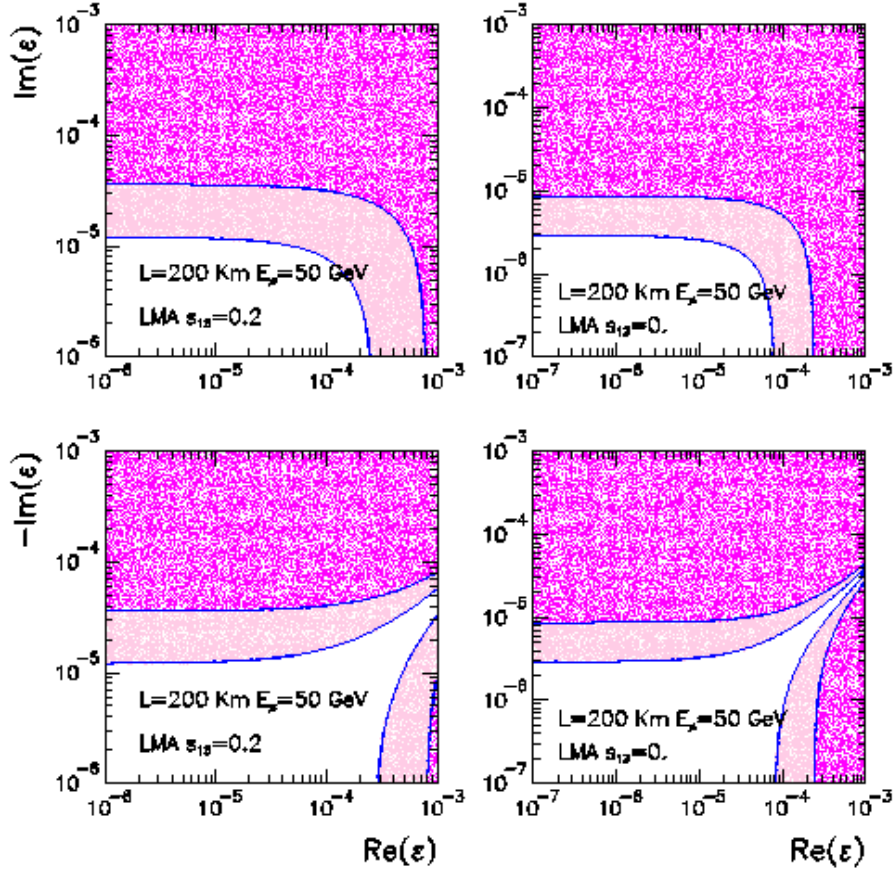}}}
\caption{Regions in the plane of [${\cal R}e(\epsilon^s_{e\mu}),
{\cal I}m(\epsilon^s_{e\mu})$] that give 
$\overline{A_{\rm CP}^{\rm NP}}/\Delta A=3$
(darker shadow) and 1 (light shadow). For the experiment, we take 
$L=200$ km, $E_\mu=50$ GeV, $10^{21}$ $\mu^-$ decays and a 40 kt detector.
For the neutrino parameters, we take $\delta=0$, 
$\Delta m^2_{31}=3\times 10^{-3}$ eV$^2$, $\tan^2\theta_{23}=1$, 
$\Delta m^2_{21}=10^{-4}$ eV$^2$ and $\tan^2\theta_{12}=1$.
In the left (right) panels we have $s_{13}=0.2(0)$. Note that the scales in the 
right panels are different from the left panels and from Fig.~\ref{cont732}.}
\label{cont200}
\end{figure}

In order to illustrate the expected improvement in sensitivity to the
New Physics when the baseline is better optimized for this particular
purpose, we plot in Fig.~\ref{cont200} the corresponding regions when
the measurement of the integrated CP asymmetry is performed at a
distance of $L=200$ km.

We would like to emphasize the following two points:

(i) Fig.~\ref{cont732} shows that $|\epsilon|$ in the range $3\times
10^{-5}$--$10^{-4}$ would lead to a ``3$\sigma$'' effect.

(ii) A shorter distance will improve the sensitivity to the new CP
violation.  Fig.~\ref{cont200} shows that, for $\delta=0$, in which
case CP--violating effects are proportional to ${\cal I}m(\epsilon)$,
an improvement by a factor of about 3 in the sensitivity to ${\cal
I}m(\epsilon)$ is expected. In contrast, the sensitivity to ${\cal
R}e(\epsilon)$ is not affected by the choice of baseline since the new
physics contribution to the matter-induced asymmetry is independent of
$L$.

(iii) A non--vanishing standard CP--violating phase, $\delta\neq0$,
together with a `large' $s_{13}$, will change the interference pattern
between the matter-induced and CP--violating contributions from new
physics. The reason is that now some of the contributions depend on
$\delta_\epsilon-\delta$, so that
${\cal R}e(\epsilon)$ and ${\cal I}m(\epsilon)$ do not correspond to 
matter-induced and CP--violating effects in any simple way.

\section{Phenomenological Constraints}
The measurements of $P_{e\mu}$ and $P_{\bar e\bar\mu}$ are sensitive
to the four effective couplings, $\epsilon^s_{e\mu}$, $\epsilon^s_{e\tau}$,
$\epsilon^d_{\mu e}$ and $\epsilon^d_{\mu\tau}$. These dimensionless couplings
represent new flavor--changing (FC) neutrino interactions. They are subject
to various phenomenological constraints. In this section, we present
these bounds in order to compare them with the experimental sensitivity
that we estimated in the previous section.

The $\epsilon^s_{e\mu}$ coupling gives the amplitude for the 
$\mu^-\to e^-\bar\nu_\mu\nu_\mu$ decay. For this process, there is no 
$SU(2)_L$-related tree-level decay that involves four charged leptons. 
Instead, by closing the 
neutrino lines into a loop, the four-Fermi coupling contributes to the 
$\mu\to e\gamma$ and $\mu\to3e$ decays. The question of how to extract reliable
bounds from loop processes in an effective theory involves many subtleties. A 
calculation in the spirit of Ref.~\cite{Burgess:1993gx}\ yields very weak 
bounds. Instead, we quote here the bound in a specific full high energy model: 
if the effective $\mu_L\overline{e_L}\nu_\mu\overline{\nu_\mu}$ coupling is 
induced by an intermediate scalar triplet, the constraint from the 
$\mu\to e\gamma$ decay reads (see, for example, \cite{CuDa})
\be
|\epsilon^s_{e\mu}|\leq5\times10^{-5}.
\label{looptri}
\ee
We emphasize again that the bound in (\ref{looptri}) is model dependent and
could be violated in models other than the one that we considered.

The $\epsilon^s_{e\tau}$ coupling gives the amplitude for the 
$\mu^-\to e^-\bar\nu_\tau\nu_\mu$ decay. The same coupling contributes also
to the $SU(2)_L$-related process $\tau^-\to\mu^+\mu^-e^-$. The experimental 
bound on the latter implies 
\be
|\epsilon^s_{e\tau}|\leq3.1\times10^{-3}.
\label{tmme}
\ee
There could be $SU(2)_L$ breaking effects that would somewhat enhance the 
neutrino couplings with respect to the corresponding charged lepton couplings. 
These effects are discussed in detail in Refs.~\cite{bg,bgp}\ where it is 
shown that they are constrained (by electroweak precision data) to
be small. Since our purpose is only to get order--of--magnitude estimates
of the bounds, we neglected the possible $SU(2)_L$ breaking effects in the
derivation of (\ref{tmme}).

The $\epsilon^d_{\mu e}$ coupling gives the amplitude for 
$\nu_e d\to\mu^- u$. It is constrained by muon conversion \cite{bg}:
\be
|\epsilon^d_{\mu e}|\lsim2.1\times10^{-6}.
\ee

The $\epsilon^d_{\mu\tau}$ coupling gives the amplitude for 
$\nu_\tau d\to\mu^- u$. It is constrained by the $\tau^-\to\mu^-\rho$ decay
\cite{bgp}:
\be
|\epsilon^d_{\mu\tau}|\lsim10^{-2}.
\ee
The bound on $|\epsilon^d_{\mu\tau}|$ is the weakest that we obtain. Moreover, 
it is not unlikely that it is indeed the largest of the couplings since it is 
the only one not to involve a first--generation lepton. For precisely the same 
reason, however, its contribution to $P_{e\mu}$ is suppressed by an additional 
power of $|U_{e3}|$, which is the reason that it is omitted in our approximate
expressions.

Let us also mention that there is a generic bound of ${\cal O}(0.1)$ on the 
purely leptonic couplings $\epsilon^s_{\alpha\beta}$ from universality in 
lepton decays and a somewhat weaker bound of ${\cal O}(0.2)$ on the 
semi-hadronic couplings $\epsilon^d_{\alpha\beta}$ from universality in pion 
decays \cite{bgp}. While universality is experimentally confirmed to high 
accuracy, these bounds are rather weak because deviations from universality 
are ${\cal O}(\epsilon^2)$.

To summarize, we expect that all the $\epsilon$'s that play a role in
the transition probabilities of interest are of ${\cal O}(10^{-3})$ or
smaller.  In the previous section, we learnt that proposed experiments
might probe these couplings down to values as small as ${\cal
O}(10^{-4})$. This means that the possibility to measure new neutrino
interactions through CP violation in neutrino oscillation experiments
is open. Conversely, such future experiments can improve the existing
bounds on FC neutrino interactions which, at present,
come from rare charged lepton decays.

\section{Conclusions and Discussion}
We summarize the main points of our study:

(i) CP--violating observables are particularly sensitive to new
physics. The reason is that the standard CP violation that comes from
the lepton mixing matrix gives effects that are particularly
suppressed by small mass differences and mixing angles. Some of these
suppression factors do not apply to new contributions.

(ii) The fact that matter effects contribute to CP-violating
observables means that these observables are sensitive to both the CP
conserving and the CP-violating contributions from New Physics.

(iii) The effects of New Physics in the production and detection
processes depend on the source--detector distance in a way that is
different from the standard one. One consequence of this situation is
that, at least in principle, it is possible to disentangle standard
and new effects. Another consequence is that in short distance
experiments the new effects are enhanced.

(iv) Our rough estimate is that future neutrino factories will be able
to probe, through CP--violating observables, effects from new
interactions that are up to about four orders of magnitude weaker than
the weak interactions.

(v) The sensitivity to New Physics effects is better than most of the
existing model--independent bounds.

We would like to mention that a similar (and, for specific models,
even stronger) level of sensitivity may be achieved by other experiments
that search for lepton flavor violation. Particularly promising are those 
involving muon decay and conversion (for a recenet review, see
\cite{kuok}): 
for example, a future experiment at PSI will be sensitive to 
$B(\mu\to e\gamma)$ at the $10^{-14}$ level \cite{barkov}, and the MECO 
collaboration has proposed an experiment to probe $\mu-e$ conversion down 
to $5\times 10^{-17}$, four orders of magnitude beyond present
sensitivities \cite{meco}. If these experiments observe a signal, the 
search for related CP violation will become of particular importance.

What type of new physics will be implied in case that a signal is observed?
The $\epsilon$ couplings represent effective four--fermion interactions
coming from the exchange of heavy particles related to New Physics.
If the New Physics takes place at some high scale $\Lambda_{\rm NP}$,
then one can set an upper bound: 
\be
\epsilon^{s,d}_{\alpha\beta}\lsim{m_Z^2\over\Lambda_{\rm NP}^2}.  
\ee
The source of this bound is in the definition of $\epsilon$, which is
the ratio of the four--fermion operator to $G_F$, and the fact that it is
maximal when the New Physics contribution comes at tree level and the
couplings are of order one. Since the expected experimental
sensitivity is to $|\epsilon|\geq{\cal O}(10^{-4})$, we learn that we
can probe models with
\be
\Lambda_{\rm NP}\lsim10\ TeV.
\ee
If the New Physics contributes to the relevant processes only at the loop
level, there is another suppression factor in $|\epsilon|$ of order
${1\over16\pi^2}$. That would mean that such models can be probed only if
$\Lambda_{\rm NP}\lsim1\ TeV$. Finally, if the flavor changing nature of
the interaction introduces a suppression factor, {\it e.g.}
$|\epsilon^s_{e\mu}|\sim m_\mu/\Lambda_{\rm NP}$, that by itself would
be enough to make it unobservable in near future experiments.
We thus learn that CP violation in neutrino oscillation experiments
will explore models with a scale that is, at most, one to two orders of
magnitude above the electroweak breaking scale, and where the flavor
structure is different from the Standard Model.

Another point concerns the Dirac structure of the four-Fermi interaction.
We did not present it explicitly in our discussion of the $G^{s,d}_{\rm NP}$
couplings. However, it is implicitly assumed in our discussion that the
Dirac structure is the same as that of the weak interactions, {\it i.e.}
a (V-A)(V-A) structure. The reason for that is that the effects that we 
discuss are a consequence of interference between weak and new interactions.
A different Dirac structure would give strong suppression factors related
to the the charged lepton masses. While our formalism would still apply,
these suppression factors would make the related effects practically
unobservable.

We conclude that a signal is likely to imply new physics at a relatively low 
scale (up to 1--10 TeV) with new sources of flavor (and, perhaps, CP) 
violation. We know of several well motivated extensions of the Standard
Model that can, in principle, induce large enough couplings. In particular,
we have in mind loop contributions involving sleptons and gauginos in
supersymmetric models, tree contributions involving charged singlet sleptons
in supersymmetric models without R-parity, and tree contributions involving
a triplet scalar in left-right symmetric models. In another class of relevant
models, such as the model of Ref.~\cite{grne}, active neutrinos mix with 
singlet neutrinos. (Here there can be $Z$-mediated contributions to the 
non-standard couplings, and the phenomenological constraints are different 
\cite{BeKa,Kita}.) A detailed analysis of new neutrino interactions within 
relevant extensions of the Standard Model is beyond the scope of this paper,
but preliminary results show that large enough couplings are allowed
and in some cases even predicted \cite{cyaynp}.

\acknowledgments
MCGG and YN thank the school of natural sciences in the Institute for
Advanced Study (Princeton), where part of this work was carried out,
for the warm hospitality.
MCG-G is supported by the European Union Marie-Curie fellowship
HPMF-CT-2000-00516.
AG is supported by Funda\c{c}\~ao Coordena\c{c}\~ao de Aperfei\c{c}oamento
de Pessoal de N\'{\i}vel Superior (CAPES).
YN is supported by the Israel Science Foundation founded by the
Israel Academy of Sciences and Humanities, by the United States - Israel
Binational Science Foundation (BSF) and by the Minerva Foundation
(Munich).
This work was also supported by the Spanish DGICYT under grants PB98-0693
and PB97-1261, by the Generalitat Valenciana under grant
GV99-3-1-01 and by the TMR network grant ERBFMRXCT960090 of the
European Union and ESF network 86. This work was also supported by the
fund for the promotion of research at the Technion.

\newpage
\appendix
\section{Transition Probability in Vacuum} 
Neglecting terms of ${\mathcal O}(\epsilon^2)$ and with no other 
approximations,  we obtain the following expression for $P_{e\mu}$:
\bea
P_{e \mu} = &&4 \sin^2 x_{21} \biggr\{ |U_{\mu 2}|^2 |U_{e 2}|^2 - {\cal R}e 
\biggr[ \; \epsilon^s_{e \mu} ( U_{e 1}^* U_{\mu 1} 
|U_{\mu 2}|^2 + U_{e 2}^* U_{\mu 2}  |U_{\mu 1}|^2 ) \no \\
  &+& \epsilon^d_{\mu e}
 (U_{\mu 2}^* U_{e 2} |U_{e 1}|^2 + U_{\mu 1}^* U_{e 1} |U_{e 2}|^2 ) 
 + \epsilon^s_{e \tau}( U_{e 2}^* U_{\mu 2} U_{\mu 1}^* U_{\tau 1}  +
 U_{e 1}^*U_{\mu 1} U_{\mu 2}^* U_{\tau 2} )  \no \\
 &+& \epsilon^d_{\mu \tau} ( U_{\mu 2}^* U_{e 2} U_{e 1}^* U_{\tau 1}+ 
U_{\mu 1}^* U_{e 1} U_{e 2}^* U_{\tau 2} ) - 
 U_{e 2}^* U_{e 3} U_{\mu 3}^* U_{\mu 2} \biggl] \biggr\} \no \\
 &+& 2 \sin 2 x_{21}\; {\cal I}m \biggr[ 
 \epsilon^s_{e \mu} ( U_{e 1}^* U_{\mu 1} |U_{\mu 2}|^2   - 
U_{e 2}^*  U_{\mu 2} |U_{\mu 1}|^2 )  \no \\
 &+& \epsilon^d_{\mu e}
 (U_{\mu 2}^* U_{e 2} |U_{e 1}|^2 - U_{\mu 1}^* U_{e 1} |U_{e 2}|^2 )
 + \epsilon^s_{e \tau}( U_{e 2}^* U_{\mu 2} U_{\mu 1}^* U_{\tau 1} - 
U_{e 1}^*U_{\mu 1}  U_{\mu 2}^* U_{\tau 2} )  \no \\
 &+& \epsilon^d_{\mu \tau} ( U_{\mu 2}^* U_{e 2} U_{e 1}^* U_{\tau 1}- 
U_{\mu 1}^* U_{e 1} U_{e 2}^* U_{\tau 2} ) +
U_{e 2}^* U_{e 3}  U_{\mu 3}^* U_{\mu 2} \biggl]   \no \\
 &+& 4 \sin^2 x_{31} \;\biggr\{|U_{\mu 3}|^2 |U_{e 3}|^2 -
  {\cal R}e \biggr[ \; \epsilon^s_{e \mu}
 (U_{e 1}^*U_{\mu 1}|U_{\mu 3}|^2 + U_{e 3}^* U_{\mu 3} |U_{\mu 1}|^2 ) \no \\
  &+& \epsilon^d_{\mu e}
 (U_{\mu 3}^* U_{e 3} |U_{e 1}|^2 + U_{\mu 1}^* U_{e 1} |U_{e 3}|^2 ) 
 + \epsilon^s_{e \tau}(U_{e 1}^* U_{\mu 1} 
U_{\mu 3}^* U_{\tau 3} + U_{e 3}^* U_{\mu 3}  U_{\mu 1}^* U_{\tau 1} ) \no \\
 &+& \epsilon^d_{\mu \tau} (U_{\mu 3}^* U_{e 3}  U_{e 1}^* U_{\tau 1} + 
U_{\mu 1}^* U_{e 1}U_{e 3}^*U_{\tau3})-
U_{e 2}^* U_{e 3}  U_{\mu 3}^* U_{\mu 2} \biggl]  \biggl\} \no \\
 &+& 2 \sin 2 x_{31}\; {\cal I}m \biggr[ \epsilon^s_{e \mu}
 (U_{e 1}^*  U_{\mu 1} |U_{\mu 3}|^2 - U_{e 3}^* U_{\mu 3} 
|U_{\mu 1}|^2 )  \no \\
 &+& \epsilon^d_{\mu e}
 (U_{\mu 3}^* U_{e 3} |U_{e 1}|^2 - U_{\mu 1}^* U_{e 1} |U_{e 3}|^2 ) 
 + \epsilon^s_{e \tau} ( U_{e 1}^* U_{\mu 1} U_{\mu 3}^* U_{\tau 3} -
 U_{e 3}^* U_{\mu 3} U_{\mu 1}^* U_{\tau 1} ) \no \\
 &+& \epsilon^d_{\mu \tau} (U_{\mu 3}^* U_{e 3}  U_{e 1}^* U_{\tau 1} - 
U_{\mu 1}^* U_{e 1} U_{e 3}^* U_{\tau 3} ) -
U_{e 2}^* U_{e 3} U_{\mu 3}^*  U_{\mu 2} \biggl] \no \\
 &-& 4 \sin^2 x_{32}\;  {\cal R}e \biggr[  \; \epsilon^s_{e \mu}
 (U_{e 2}^*  U_{\mu 2} |U_{\mu 3}|^2 + U_{e 3}^* U_{\mu 3}|U_{\mu2}|^2 ) \no \\
  &+& \epsilon^d_{\mu e}
 (U_{\mu 3}^* U_{e 3} |U_{e 2}|^2 + U_{\mu 2}^* U_{e 2} |U_{e 3}|^2 ) 
 + \epsilon^s_{e \tau}( U_{e 2}^* U_{\mu 2} U_{\mu 3}^* U_{\tau 3} +
 U_{e 3}^* U_{\mu 3} U_{\mu 2}^* U_{\tau 2} ) \no \\
 &+& \epsilon^d_{\mu \tau} ( U_{\mu 3}^* U_{e 3} U_{e 2}^* U_{\tau 2}  +
 U_{\mu 2}^* U_{e 2} U_{e 3}^* U_{\tau 3}  ) + 
 U_{e 2}^* U_{e 3} U_{\mu 3}^*  U_{\mu 2} \biggl]  \no \\
 &+& 2 \sin 2 x_{32} {\cal I}m \biggr[ \epsilon^s_{e \mu} 
(U_{e 2}^* U_{\mu 2} |U_{\mu 3}|^2-U_{e 3}^* U_{\mu 3} |U_{\mu 2}|^2 )\no \\
 &+&\epsilon^d_{\mu e} (U_{\mu 3}^* U_{e 3} |U_{e 2}|^2 -
  U_{\mu 2}^* U_{e 2} |U_{e 3}|^2 ) 
 +\epsilon^s_{e \tau} ( U_{e 2}^* U_{\mu 2} U_{\mu 3}^* 
U_{\tau 3} - U_{e 3}^*  U_{\mu 3} U_{\mu 2}^* U_{\tau 2} ) \no \\
 &+&\epsilon^d_{\mu \tau} (U_{\mu 3}^* U_{e 3} U_{e 2}^* U_{\tau 2} - 
 U_{\mu 2}^* U_{e 2} U_{e 3}^* U_{\tau 3} ) +
  U_{e 2}^* U_{e 3} U_{\mu 3}^*  U_{\mu 2} \biggl].
\label{pemuvac}
\eea


\newpage
%
%
\end{document}